\documentclass[aps,prc,twocolumn,nofootinbib,amsmath,showpacs,superscriptaddress,floatfix]{revtex4}
\usepackage{graphicx,epsfig,epstopdf,longtable}
\usepackage{mathptmx}
\usepackage[light,first]{draftcopy} 
\usepackage[pdftex]{hyperref}
\usepackage{wasysym}

\newcommand{\beqy}{\begin{eqnarray}}
\newcommand{\eeqy}{\end{eqnarray}}
\newcommand{\bmlet}{\begin{subequations}}
\newcommand{\emlet}{\end{subequations}}
\newcounter{saveeqn}

\def\gsimeq{\,\,\raise0.14em\hbox{$>$}\kern-0.76em\lower0.28em\hbox  
{$\sim$}\,\,}  
\def\lsimeq{\,\,\raise0.14em\hbox{$<$}\kern-0.76em\lower0.28em\hbox  
{$\sim$}\,\,}  

\begin{document}

\title{Level density and $\gamma$-ray strength function in the odd-odd $^{238}$Np nucleus}
\author{T.G.~Tornyi}
\email{tornyitom@gmail.com}
\affiliation{Department of Physics, University of Oslo, N-0316 Oslo, Norway}
\affiliation{Institute of Nuclear Research of the Hungarian Academy of Sciences (MTA Atomki), Debrecen, Hungary}
\author{M.~Guttormsen}
\affiliation{Department of Physics, University of Oslo, N-0316 Oslo, Norway}
\author{T.K.~Eriksen}
\affiliation{Department of Physics, University of Oslo, N-0316 Oslo, Norway}
\author{A.~G{\"o}rgen}
\affiliation{Department of Physics, University of Oslo, N-0316 Oslo, Norway}
\author{F.~Giacoppo}
\affiliation{Department of Physics, University of Oslo, N-0316 Oslo, Norway}
\author{T.W.~Hagen}
\affiliation{Department of Physics, University of Oslo, N-0316 Oslo, Norway}
\author{A.~Krasznahorkay}
\affiliation{Institute of Nuclear Research of the Hungarian Academy of Sciences (MTA Atomki), Debrecen, Hungary}
\author{A.C.~Larsen}
\affiliation{Department of Physics, University of Oslo, N-0316 Oslo, Norway}
\author{T.~Renstr{\o}m}
\affiliation{Department of Physics, University of Oslo, N-0316 Oslo, Norway}
\author{S.J.~Rose}
\affiliation{Department of Physics, University of Oslo, N-0316 Oslo, Norway}
\author{S.~Siem}
\affiliation{Department of Physics, University of Oslo, N-0316 Oslo, Norway}
\author{G.M.~Tveten}
\affiliation{Department of Physics, University of Oslo, N-0316 Oslo, Norway}

\date{\today}

\begin{abstract}
The level density and $\gamma$-ray strength function in the quasi-continuum of $^{238}$Np
have been measured using the Oslo method.
The level density function follows closely the constant-temperature level density formula
and reaches 43 million levels per MeV at $S_n=5.488$ MeV of excitation energy.
The $\gamma$-ray strength function displays a two-humped resonance at low-energy as also seen in previous
investigations of Th, Pa and U isotopes. The structure is interpreted as the scissors resonance and has
an average centroid of $\omega_{\rm SR}=2.26(5)$~MeV and a total strength of $B_{\rm SR} = 10.8(12) \mu_{N}^{2}$, which is in excellent agreement with sum-rule estimates.
The scissors resonance is shown to have an impact on the $^{237}$Np$(n, \gamma)^{238}$Np cross section.

\end{abstract}


\maketitle

\section{Introduction}
\label{sec:int}
Atomic nuclei in the actinide region are believed to be
synthesized in explosive stellar environments purely by the rapid neutron-capture process.
Therefore, to predict their abundances found on Earth~\cite{ar07,kaeppler2011},
one has to know the various reaction rates for all isotopes including the ones with extreme neutron excess.
Reaction rates are also vital for the modeling of future and existing nuclear reactors~\cite{aliberti2006,chadwick2011}.
It is particularly important to ensure a reliable extrapolation in cases where measured data are insufficient or lacking.

The $^{237}$Np isotope with a half-life of 2.14 million years is one of the main constituents in nuclear spent fuel.
In the former US high-level waste repository in the Yucca Mountain, Nevada, about 40 tons of $^{237}$Np are stored~\cite{esch2008},
and it is of great interest to find methods for transmuting this type of radioactive waste.
In order to obtain high transmutation efficiency, the neutron fission-to-capture ratio should be determined for the particular isotope
as function of neutron energy. Hence, accurate fission and capture cross sections are necessary to make reliable predictions~\cite{aliberti2004}.

The nuclear level density and $\gamma$-ray strength function ($\gamma$SF) are important inputs in statistical
Hauser-Feshbach reaction-rate calculations. These functions describe the average properties of excited
nuclei in the quasi-continuum region, where the number of levels is too high to study individual 
states and their transitions. Here, the Oslo  method~\cite{Schiller00,Lars11} has been shown to be an excellent tool
to determine simultaneously the level density and the $\gamma$-ray strength function ($\gamma$SF).

Recently, the Oslo method
was applied to the $^{231-233}$Th, $^{232,233}$Pa and $^{237-239}$U isotopes~\cite{guttormsen2012,nld2013,gsf2014}.
The level densities of all eight actinides follow closely the constant-temperature
level density formula. Furthermore, a large scissors resonance (SR) was observed in
the $\gamma$SF with a $\gamma$-energy centroid at $\omega_{\rm SR} \approx 2.4$ MeV. This extra $\gamma$
strength enhances the decay with $\gamma$ rays  relative to other decay branches such as particle emission or fission.

One would expect that the SR is present throughout the region of well-deformed actinides. The n\_TOF
collaboration~\cite{ntof2011} has recently reported on $(n, \gamma)$ experiments on the $^{234}$U, $^{237}$Np and $^{240}$Pu isotopes.
They verify a low-energy structure in
$^{235}$U and $^{241}$Pu, but not in $^{238}$Np, a result which
is rather surprising. The odd-odd $^{238}$Np nucleus has the same gross properties as other actinides,
and the Oslo group has confirmed that the structure also appears in the odd-odd $^{232}$Pa nucleus~\cite{gsf2014}.
Thus, the n\_TOF results on $^{238}$Np have triggered us to investigate this case further.

The main purpose of the present work is to search for the SR in $^{238}$Np
and to determine the total level density and $\gamma$SF. Furthermore, we present for the first
time $(n, \gamma$) cross-section from Hauser-Feshbach calculations using
the measured level density and $\gamma$SF as inputs. The calculations are compared with known $(n, \gamma$) data from literature.

The manuscript is organized as follows. Section II describes briefly the experimental methods,
and in Sect.~III the extraction and normalization of the level density and $\gamma$SF are discussed.
In Sect.~IV the SR is presented, and extracted resonance parameters are compared 
to previous results and sum-rules estimates. In Sect.~V the measured level density and $\gamma$SF
are used as inputs to Hauser-Feshbach calculations in order to estimate $(n, \gamma)$ cross sections.
Conclusions are drawn in Sect.~VI.

\section{Experiment}
\label{sec:exp}

The experiment was performed with the MC-35
Scanditronix cyclotron at the Oslo Cyclotron Laboratory (OCL). 
The $^{237}$Np target (thickness 0.200 mg/cm$^2$ and enrichment 99\%), which had
a carbon backing (thickness 0.020 mg/cm$^2$), was bombarded with a 13.5 MeV deuteron beam.
Particle-$\gamma$ coincidences were measured with the SiRi particle telescope and the 
CACTUS $\gamma$-detector system~\cite{siri,CACTUS}.
 \begin{figure*}[t]
 \begin{center}
 \includegraphics[clip,width=2\columnwidth]{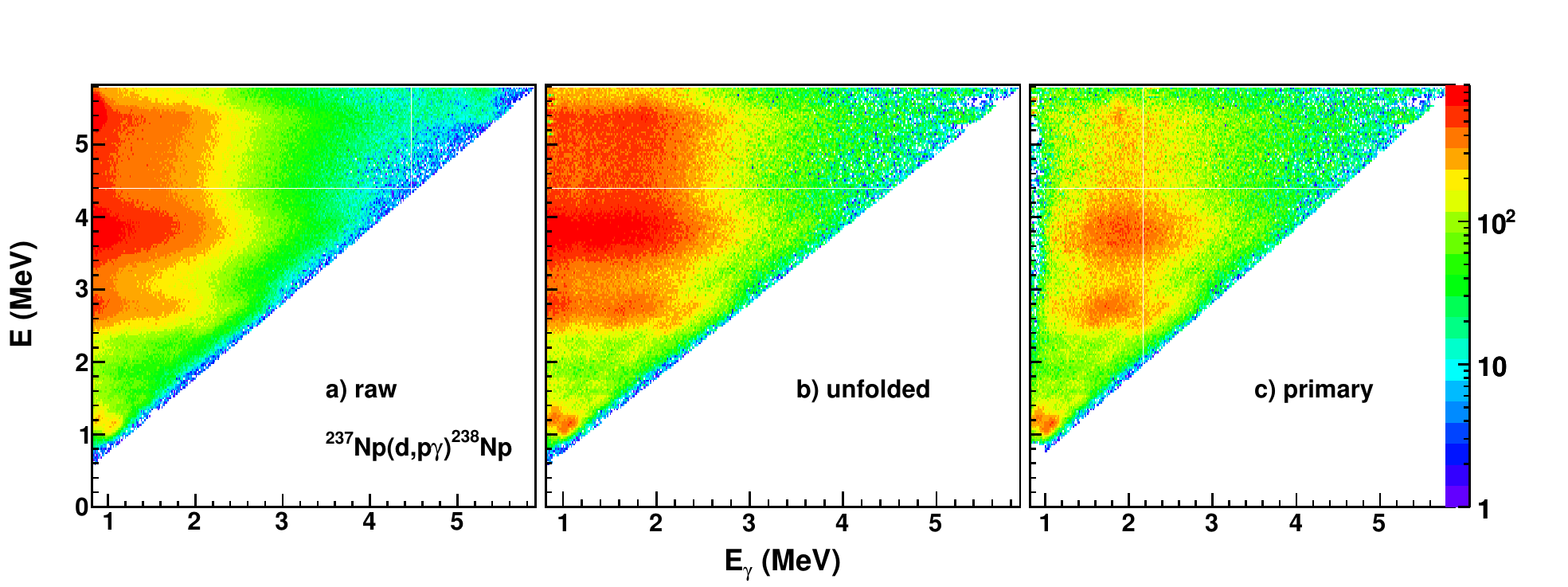}
 \caption{(Color online) Initial excitation energy $E$ versus $\gamma$-ray energy 
 $E_{\gamma}$ from particle-$\gamma$ coincidences recorded with the 
 $^{237}$Np$(d,p\gamma)^{238}$Np reaction. The raw $\gamma$-ray spectra
 (a) are first unfolded  by the NaI response function (b) and finally
  the primary or first-generation
 $\gamma$-ray spectra (c) are extracted as function of excitation energy $E$.
 The excitation  and $\gamma$ energy  axis have dispersions of
 14.0~keV/ch and 30.4~keV/ch, respectively.}
 \label{fig:matrix}
 \end{center}
 \end{figure*}

The 64 SiRi telescopes were placed in backward direction
covering eight angles from $\theta = 126^\circ$ to $140^\circ$
relative to the beam axis. This configuration was chosen to reduce the intense elastically scattered deuterons
and to obtain a broad and rather high  spin distribution
that matches better to the spin distribution of available states in the quasi-continuum. The front
and back detectors have thicknesses of $130$~$\mu$m and $1550$~$\mu$m, respectively. 
The CACTUS array consists of 28 collimated $5^{\prime\prime} \times 5^{\prime\prime}$ NaI(Tl) 
detectors with a total efficiency of $15.2$\% at $E_{\gamma} = 1.33$~MeV. 

The $E$ back detectors were used as master gates and the start for the time-to-digital-converter (TDC). One or more
of the NaI detectors were used as individual TDC stops. In this way, prompt
particle-$\gamma$ coincidences with background subtraction could be sorted event by event.
The proton events were selected by setting proper 2-dimensional gates on the 64 $\Delta$E-E matrices.
From the kinematics of the reaction, the proton energies deposited in the telescopes
were translated into initial excitation energy $E$ in the residual $^{238}$Np nucleus.

Figure~\ref{fig:matrix} shows the first main steps of the Oslo method.
After sorting the data into a raw matrix of initial excitation energy versus the NaI energy signal (a), the
matrix is unfolded~\cite{Gutt96} using the NaI response function for each excitation bin (b). In
panel (c) the first-generation (primary) $\gamma$-ray matrix $P(E,E_{\gamma})$ is shown.
Here, an iterative subtraction technique was applied to separate out the distribution of the
first-generation $\gamma$s from the total $\gamma$
cascade~\cite{Gutt87}. The technique is based on the assumption that the $\gamma$ distribution
is the same whether the levels were populated directly
by the nuclear reaction or by $\gamma$ decay from higher-lying states. 
This assumption is necessarily fulfilled when states have the same
relative probability to be populated by the two processes, since $\gamma$-branching 
ratios are properties of the levels themselves.

The first generation matrix $P$ is built from the total matrix $P_{{\rm gen} > 0}$ of Fig.~\ref{fig:matrix} (b),
where all $\gamma$s of all cascade are included. The matrix with higher generations
$P_{{\rm gen} > 1}$ is obtained by weighting and summing
the spectra at lower excitation energy. In principle, the first-generation matrix $P_{{\rm gen} = 1}$ is identical to the proper weighting function and obtained by an iterative procedure described in detail in Ref.~\cite{Gutt87}.

The number of counts in the second or higher-generation spectra $A_{{\rm gen} > 1}$ has to relate
to the counts of the total spectrum $A_{{\rm gen} > 0}$.
Since the $\gamma$ multiplicity of the first-generation spectra equals unity, we find
\begin{equation}
A_{{\rm gen} > 1}=\frac{M_{\gamma}(E)-1}{M_{\gamma}(E)}A_{{\rm gen} > 0}.
\end{equation}
Provided, that we have a correct normalization of the counts in the $P_{{\rm gen} > 1 }$ matrix, the primary matrix
is given by
$P=P_{{\rm gen} > 0} - P_{{\rm gen} > 1}$.
The average $\gamma$ multiplicity from initial excitation energy $E$ is given by
\begin{equation}
M_{\gamma}(E)=\frac{E}{\langle E_{\gamma}(E)\rangle},
\end{equation}
where  $\langle E_{\gamma}(E)\rangle$ is the centroid of the total $\gamma$ spectrum [Fig.~\ref{fig:matrix} (b)]
at $E$.
 \begin{figure}[h]
 \begin{center}
 \includegraphics[clip,width=\columnwidth]{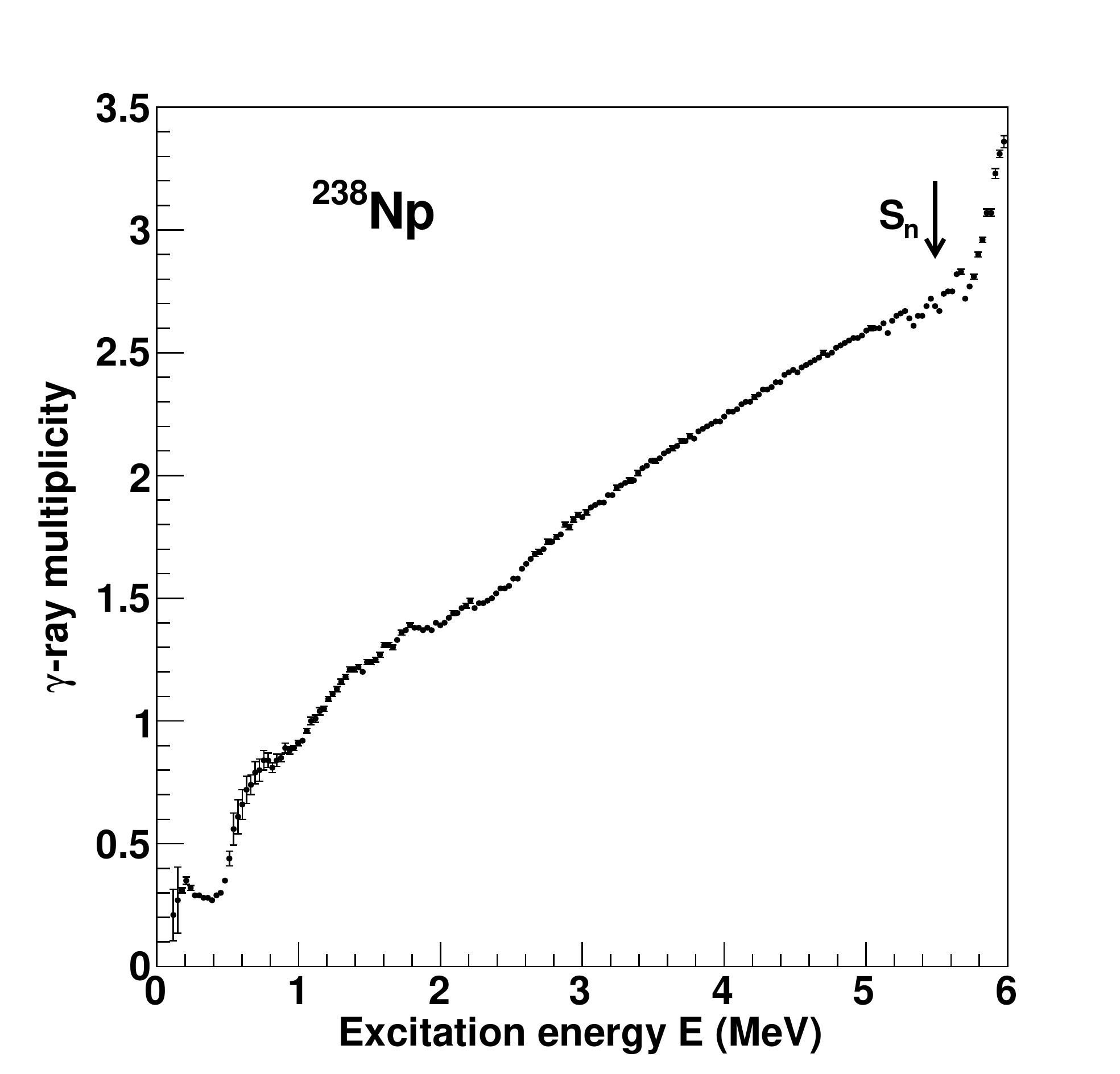}
 \caption{Gamma-ray multiplicity for $E_{\gamma} > 0.45$ MeV as function of excitation energy $E$ in $^{238}$Np.}
 \label{fig:mult}
 \end{center}
 \end{figure}

Figure~\ref{fig:mult} shows the $\gamma$ multiplicity for $E_{\gamma} > 0.45$ MeV 
as function of initial excitation energy $E$.
At the lower excitation energies, the multiplicity is seen to fluctuate since
the decay routes become increasingly dependent on available levels of certain
spin/parity and structure when approaching the ground state. Above $E=2-3$ MeV, the decay
seems to reveal a statistical behavior. To proceed with the Oslo method, we use only the region
$E=3.0-5.7$ MeV of the first generation matrix of Fig.~\ref{fig:matrix} (c).

 \begin{figure}[t]
 \begin{center}
 \includegraphics[clip,width=\columnwidth]{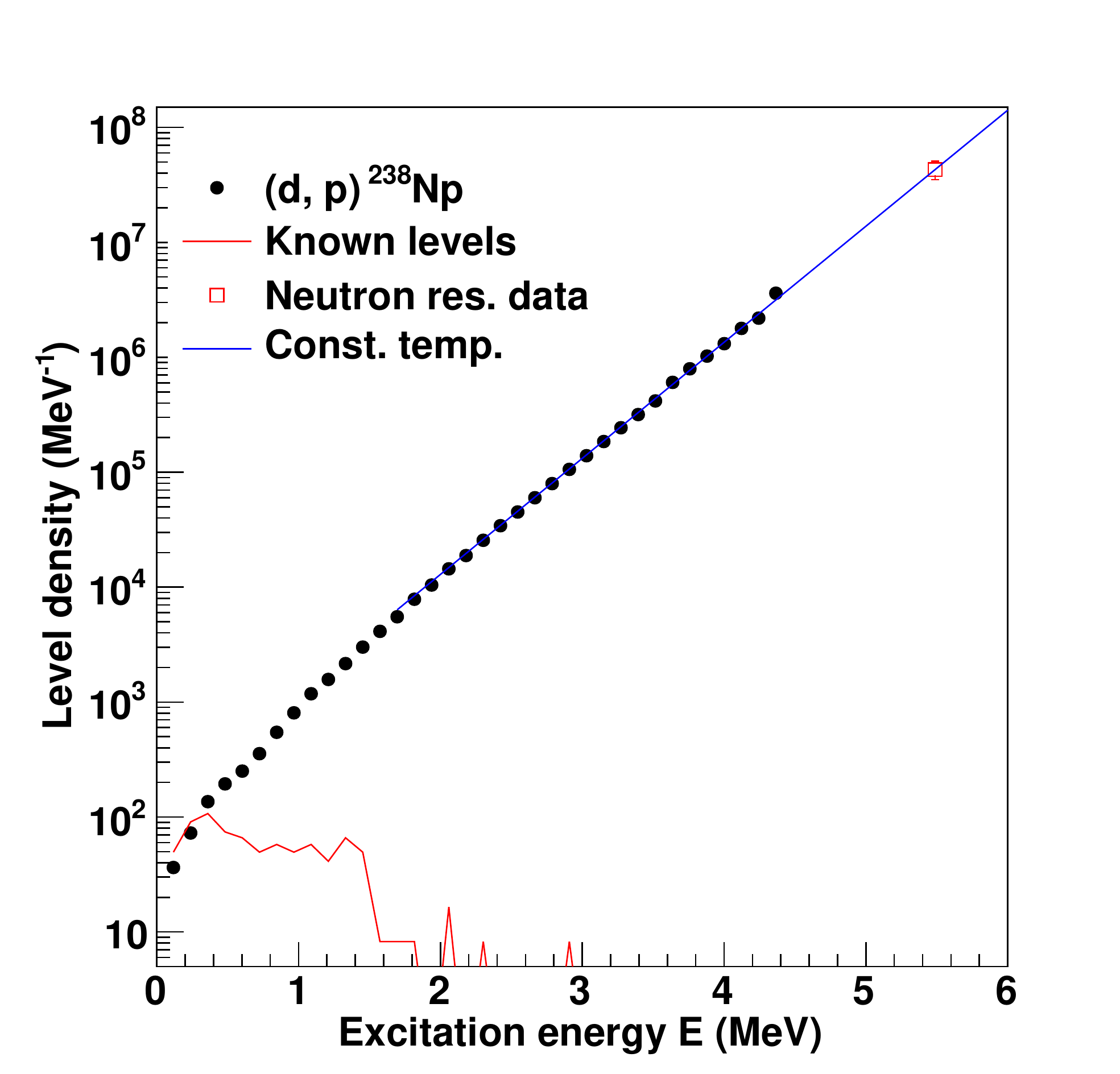}
 \caption{(Color online) Level density for $^{238}$Np. The experimental data are normalized
 to the level density of known discrete levels at low excitation energy $E$ (red solid line) and
 the level density extracted from known neutron resonance spacings $D_0$ at the neutron separation energy $S_n$.
 The connection between $\rho(S_n)$ (the upper right data points) and our experimental data 
 are made with a constant-temperature formula with $T_{\rm CT}= 0.43$ MeV.
 The odd-odd $^{238}$Np nucleus has an extreme
 high level density of $\approx$ 43 million levels per MeV at the neutron separation energy of $S_n=5.488$ MeV.}
 \label{fig:rhotot}
 \end{center}
 \end{figure}

According to the
Brink hypothesis~\cite{brink}, the $\gamma$-ray transmission coefficient ${\cal {T}}$ is 
approximately independent of excitation energy. Thus, the first-generation
matrix $P(E,E_{\gamma})$ may be factorized as follows:
\begin{equation}
P(E, E_{\gamma}) \propto   {\cal{T}}(E_{\gamma}) \rho (E -E_{\gamma}),\
\label{eqn:3}
\end{equation}
where $\rho (E -E_{\gamma})$ is the level density at the 
excitation energy after the first $\gamma$-ray has been emitted in the cascades.
This factorization allows the disentanglement  of the level density and $\gamma$-ray transmission coefficient. Note that no initial assumptions are made regarding to the functional form of ${\cal{T}}$ and $\rho$.
However, the least-square fit of ${\cal{T}}\rho$ to the measured matrix $P$ [see Eq.~(\ref{eqn:3})]
determines only the functional form of ${\cal{T}}$ and $\rho$;
if one solution of the functions ${\cal{T}}$ and $\rho$ is known,
one may construct infinitely many identical fits to the
$P(E,E_{\gamma})$ matrix by
\begin{eqnarray}
\tilde{\rho}(E-E_\gamma)&=&A\exp[\alpha(E-E_\gamma)]\,\rho(E-E_\gamma),
\label{eq:array1}\\
\tilde{{\mathcal{T}}}(E_\gamma)&=&B\exp(\alpha E_\gamma){\mathcal{T}} (E_\gamma).
\label{eq:array2}
\end{eqnarray}
The transformation parameters $A$, $\alpha$ and $B$ have then to be determined from other data, 
which is discussed in the next section.

    \begin{table*}[]
    \caption{Parameters used to extract level density and $\gamma$SF (see text).}
    \begin{tabular}{ccccc|cc|c}
    \hline
    \hline
    $S_n$ & a & $E_1$   &  $\sigma(S_n)$&  $D_0$   &   $\rho(S_n)$ &  $\rho(S_n)_{\rm red}$    &$\langle \Gamma_{\gamma}(S_n)\rangle$  \\
    (MeV)&(MeV$^{-1})$ & (MeV)&      & (eV)   &(10$^6$MeV$^{-1}$)&(10$^6$MeV$^{-1}$)& (meV)       \\
    \hline
    5.488 & 25.96     &-0.84 & 8.28&0.57(3) &    43.0(78)      &    22            &   40.8(12)      \\
    \hline
    \hline
    \end{tabular}
    \label{tab:parameters}
    \end{table*}

\section{Normalization}
We need to find the $A$ and $\alpha$ parameters of Eq.~(\ref{eq:array1}) in order
to determine the level density. The two normalization points are determined 
at low excitation energy from the known level scheme~\cite{ENSDF}
and at high energy from the density of neutron resonances following thermal ($n$, $\gamma$) capture
at the neutron separation energy $S_n$.
Here, the upper data point $\rho(S_n)$
is estimated from $\ell = 0$ neutron resonance spacings $D_0$ taken from RIPL-3~\cite{RIPL3}
assuming a spin distribution~\cite{GC}
\begin{equation}
g(E=S_n,I) \simeq \frac{2I+1}{2\sigma^2}\exp\left[-(I+1/2)^2/2\sigma^2\right].
\label{eq:spindist}
\end{equation}
The spin-cutoff parameter was determined from the global systematic study of level-density parameters by von Egidy and Bucurescu,
who use a rigid-body moment of inertia approach~\cite{egidy2}:
\begin{equation}
\sigma^2 = 0.0146 A^{5/3} \frac{1+\sqrt{1+4aU}}{2a},
\label{eqn:8}
\end{equation}
where $A$ is the mass number, $a$ is the level density parameter, $U=E-E_1$ is the intrinsic excitation energy, and $E_1$ is the back-shift parameter.
Table~\ref{tab:parameters} lists the  $D_0$,  $\sigma$ and $\rho$ values at $S_n$ used to determine
the level density. The $a$ and $E_1$ parameters are taken from Ref.~\cite{egidy2}. 
One should note that the spin distribution at such high excitation energies is not well known, 
and thus imposes a systematic uncertainty on our results.

Figure \ref{fig:rhotot} demonstrates
how the level density is normalized to the anchor points at low and high excitation energies. 
The level density follows closely the constant temperature formula with $\ln \rho \propto E/T_{\rm CT}$ as also measured for other Th, Pa and U isotopes~\cite{nld2013}.
It is interesting to see that only a small fraction of the levels, even at low excitation energies, have been
observed in the odd-odd $^{238}$Np. The reason is of course the very high level density,
 e.g.~at 1  MeV of excitation energy the average distance between levels is  $\approx 1$ keV, only.
    
The level density is closely related to the entropy of the system, from which thermodynamic
quantities such as temperature and heat capacity can be extracted. This
will not be further elaborated here since the properties of the level density function observed for $^{238}$Np are very similar to those observed for $^{237-239}$U ~\cite{nld2013}.

The light-ion $(d,p)$ reaction used in this work may not populate the highest spins levels available in the nucleus, which in turn
could influence the shape of the observed primary $\gamma$ spectra $P$. Since the transmission coefficient
${\cal{T}}$ is assumed to be independent of spin, the observed $P$ matrix should be 
fitted with the product ${\cal{T}}\rho_{\rm red}$, where the reduced level density
is extracted by assuming a lower value of $\rho$ at $S_n$. Since there are uncertainties
in the total $\rho(S_n)$ through the estimate of $\sigma$ and also the actual spin distribution 
brought into the nuclear system by the specific reaction, the extracted slope of ${\cal{T}}$
becomes rather uncertain.

The parameter $B$ controls the scaling of the transmission coefficient ${\cal {T}}(E_{\gamma})$. 
Here we use the average, total radiative width $\langle \Gamma_{\gamma} \rangle$ at $S_n$ assuming 
 that the $\gamma$-decay is dominated by dipole transitions. For initial spin $I$ and parity $\pi$, the width is given by~\cite{ko90} 
\begin{eqnarray}
\langle\Gamma_\gamma\rangle=\frac{1}{2\pi\rho(S_n, I, \pi)} \sum_{I_f}&&\int_0^{S_n}{\mathrm{d}}E_{\gamma} B{\mathcal{T}}(E_{\gamma})
\nonumber\\
&&\times \rho(S_n-E_{\gamma}, I_f),
\label{eq:norm}
\end{eqnarray}
where the summation and integration run over all final levels with spin $I_f$ that are accessible by $E1$ or $M1$
transitions with energy $E_{\gamma}$. 

Since our spin distribution for the reaction is likely to be lower than 
the spin distribution of the available levels,  the standard normalization procedure of the Oslo method~\cite{Schiller00,voin1} to
determine the $\alpha$ parameter for the transmission coefficient in Eq.~(\ref{eq:array2}) is not reliable.
Instead we compare
the $\gamma$SF with the extrapolation of known data from photo-nuclear reactions.
 \begin{figure}[t]
 \begin{center}
 \includegraphics[clip,width=\columnwidth]{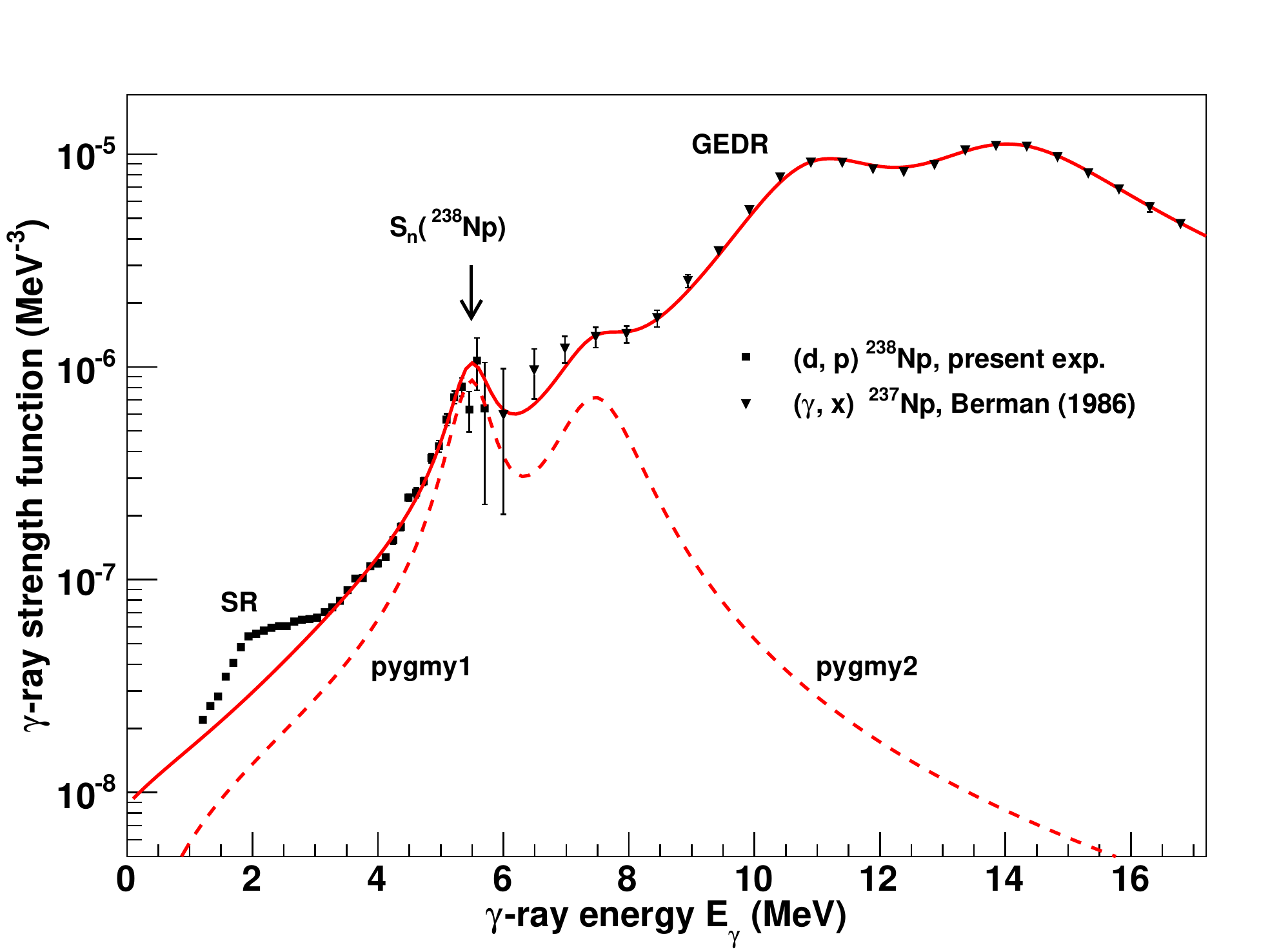}
 \caption{(Color online) Experimental $\gamma$SF from the present $(d,p)^{238}$Np experiment (black filled squares) compared
 with the estimated underlying $\gamma$SF (red curve), which represents the strength expected without the SR. The
 ($\gamma$, x) data (black filled triangles) are taken from Berman {\em et al.}~\cite{berman1986}.}
 \label{fig:GDR}
 \end{center}
 \end{figure}
\begin{table*}[]
\caption{Resonance parameters used for the $\gamma$SF extrapolation.}
\begin{tabular}{ccc|ccc|c|ccc|ccc}
\hline
\hline
$\omega_{E1,1}$&$\sigma_{E1,1}$&$\Gamma_{E1,1}$&$\omega_{E1,2}$&$\sigma_{E1,2}$&$\Gamma_{E1,2}$&$T_f$&$\omega_{\rm pyg1}$&$\sigma_{\rm pyg1}$&$\Gamma_{\rm pyg1}$&$\omega_{\rm pyg2}$&$\sigma_{\rm pyg2}$&$\Gamma_{\rm pyg2}$ \\
 (MeV)  &      (mb)     &      (MeV)    &   (MeV)  &     (mb)      &    (MeV)      &(MeV)&    (MeV)    &      (mb)        &          (MeV)    &(MeV)  &     (mb)    &  (MeV)    \\ \hline
11.3    &      970      &       3.0     &     14.6 &      1520      &     4.4       & 0.2 &     5.5    &      50          &        0.7        & 7.5 &     60    &    1.4   \\

\hline
\end{tabular}
\\
\label{tab:GDR_param}
\end{table*}

The  $\gamma$SF for dipole radiation can be calculated from the transmission coefficient ${\cal {T}}(E_{\gamma})$ by~\cite{RIPL3}
\begin{equation}
f (E_{\gamma}) =\frac{1}{2\pi} \frac{ {\mathcal{T}}(E_{\gamma})}{ E_{\gamma}^3}.
\label{eq:fT}
\end{equation}
These data are compared with the strength function derived from the cross section $\sigma$
of  photo-nuclear reactions by~\cite{RIPL3}
\begin{equation}
f (E_{\gamma}) =\frac{1}{3\pi^2 \hbar^2c^2} \frac{\sigma(E_{\gamma})}{ E_{\gamma}},
\label{eq:fT2}
\end{equation}
where the factor $1/3\pi^2\hbar^2c^2$ takes the value $8.6737\times10^{-8}$~mb$^{-1}$MeV$^{-2}$.
In Fig.~\ref{fig:GDR} the $\gamma$SF derived from $^{237}$Np($\gamma$, x) cross section
  by Berman {\em et al.}~\cite{berman1986} is shown (x means all possible ejectiles, as well as fission fragments).
We assume that this strength do not vary much from $^{237}$Np to $^{238}$Np, as pointed out for the two
 $^{236,238}$U isotopes~\cite{gsf2014}.

Since our data cover $\gamma$ energies below $S_n$, we have to extrapolate the ($\gamma$, x) data to lower energies.
For the double-humped giant electric dipole resonance (GEDR) we fit the data with two enhanced generalized Lorentzians
(EGLO) as defined in RIPL~\cite{RIPL3}, but with a constant temperature parameter of the final states $T_f$, in accordance with the Brink hypothesis. 
In addition the ($\gamma$, x) data~\cite{berman1986} reveal a knee at around 7.5 MeV indicating a resonance-like structure (labeled pygmy2 in Fig.~\ref{fig:GDR}).
We also note the steep flank of our $\gamma$SF data from 4 to 5 MeV of $\gamma$ energy. In order to match this increase in the $\gamma$SF another pygmy
is postulated at around 5.5 MeV. The two pygmy resonances are described by simple Lorentzians:
\begin{equation}
f_{\rm pyg}=\frac{1}{3\pi^2\hbar^2c^2}\frac{\sigma_{\rm pyg}\Gamma_{\rm pyg}^2E_{\gamma}}
{(E_{\gamma}^2 - \omega_{\rm pyg}^2)^2+ \Gamma_{\rm pyg}^2E_{\gamma}^2}.
\end{equation}
The sum of the two GEDR and the two pygmy $\gamma$SFs are shown as a solid red curve in Fig.~\ref{fig:GDR}.
The four sets of resonance parameters are listed in Table~\ref{tab:GDR_param}.
    
We have also tested another approach of modeling the $\gamma$SF in the 4 - 8 MeV region. One broad Gaussian 
shape at 6.5~MeV gives approximatelly the same fit to the available data. 
However, we feel that there are no arguments to adopt a Gaussian 
shape for a resonance structure. Since the choice of one broad Lorentzian fails to reproduce the data,
we keep to the assumption of two narrow pygmys as shown in Fig.~\ref{fig:GDR}.

Provided that the extrapolation in Fig.~\ref{fig:GDR} (red solid curve) is reliable,
we may assume that this $\gamma$SF represents the "base line" with no additional strength from other resonances.
Thus, we normalize the measured $\gamma$SF to this underlying background.
Here, the $\alpha$ parameter is adjusted to obtain the right slope of the observed $\gamma$SF; the
level density at $S_n$ had to be reduced from 43 to 22 million levels per MeV. The $B$ parameter
was determined by use of Eq.~(\ref{eq:norm}) in order to reproduce the experimental $\gamma$ width $\langle\Gamma_\gamma\rangle$
listed in Table~\ref{tab:parameters}.

\section{The scissors resonance}
\begin{figure}[t]
 \begin{center}
 \includegraphics[clip,width=\columnwidth]{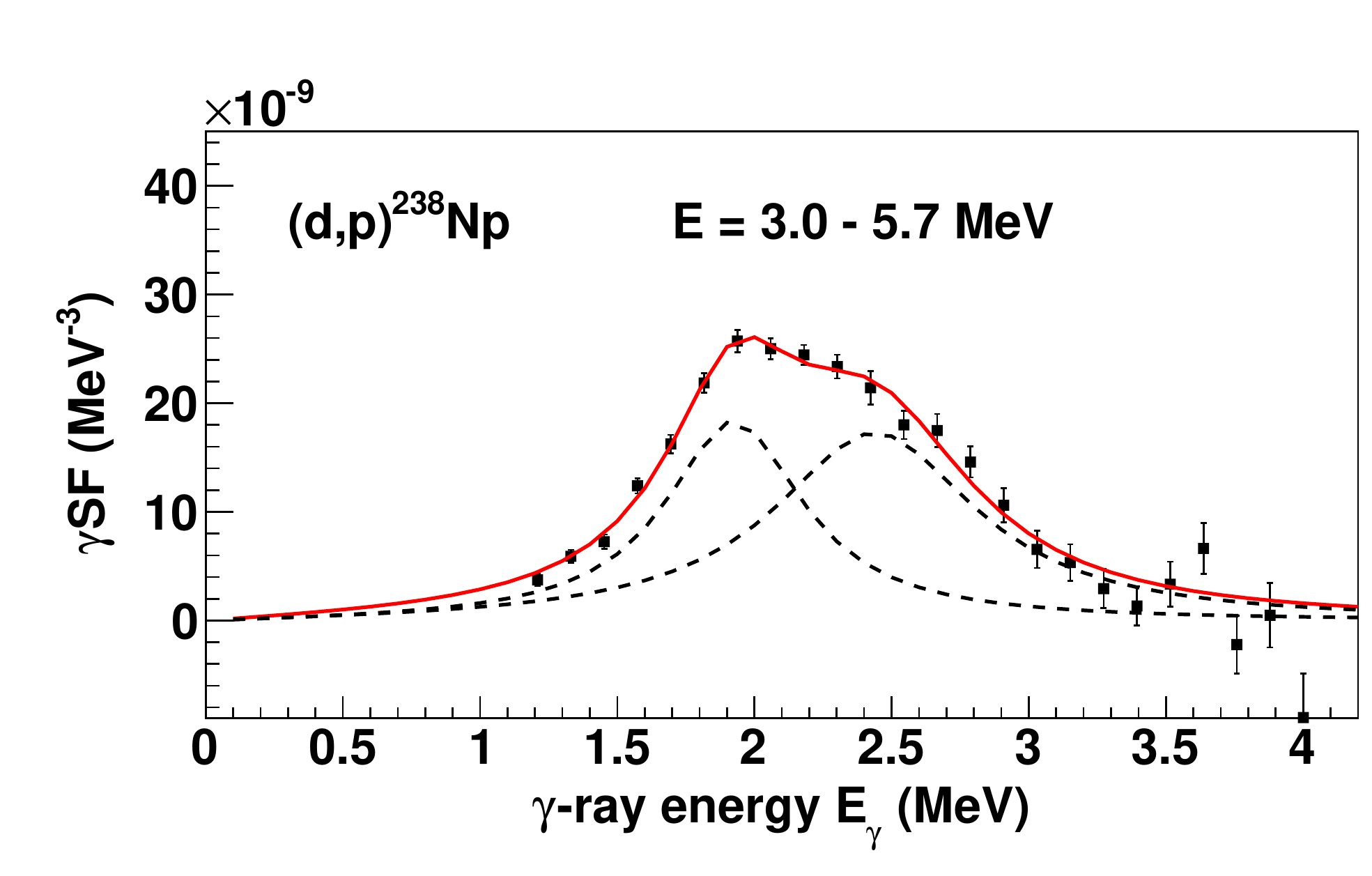}
 \caption{(Color online) The extracted $\gamma$SF for the scissors resonance in the quasi-continuum of $^{238}$Np.}
 \label{fig:pygmy}
 \end{center}
 \end{figure}

Figure~\ref{fig:pygmy} shows the $\gamma$SF where the assumed Lorentzian shape line of Fig.~\ref{fig:GDR}
has been subtracted. The observed structure, which is interpreted as the SR, is in accordance with 
previous observations in the $^{231-233}$Th, $^{232,233}$Pa and $^{237-239}$U isotopes~\cite{guttormsen2012,nld2013,gsf2014}.
Thus, our findings is in strong disagreement
with the $(n,\gamma)^{238}$Np results of the n\_TOF group that found no evidence for the SR structure~\cite{ntof2011}.
    
The SR is split into two components where the strengths of each component is given by a set of resonance parameters:
\begin{equation}
    B=\frac{9\hbar c}{32 \pi ^2}\left( \frac{\sigma \Gamma}{\omega}\right).
\end{equation} The resonance parameters of the lower and upper component, as well as the total strength
and average energy centroid are listed in Table~\ref{tab:strengths}.

We find that the separation in energy between the two components is much smaller than
previously seen for Th, Pa and U~\cite{gsf2014}; 
$\Delta\omega_{\rm SR}=0.89(15)$ compared to $0.53(6)$ MeV for $^{238}$Np.
In addition the higher lying component takes the main strength contrary
to the other actinides where the low lying strength carried almost 2/3 of the strength.
The total strength is the same as for the other actinides within the uncertainties.

Recent high quality measurements at the High-Intensity $\gamma$-ray Source (HI$\gamma$S)
at the Triangle Universities Nuclear Laboratory (TUNL) has discovered more strength than for previous
($\gamma, \gamma '$) measurements in this mass region~\cite{heil1988,margraf1990,yevetska2010}.
In $^{232}$Th a strength
of $B_{\rm SR}=4.3(6)\mu_N^2$ at $\omega_{\rm SR}=2.5(4)$ MeV has been reported~\cite{adekola2011} and for
$^{238}$U there has been measured $B_{\rm SR}=8(1)\mu_N^2$ at $\omega_{\rm SR}=2.6(6)$ MeV~\cite{hammond2012}.

\begin{table*}[]
\caption{Scissors resonance parameters of $^{238}$Np and its sum-rule estimates [Eqs.~(\ref{eq:omega}) and (\ref{eq:b}), see text].}
\begin{tabular}{c|cccc|cccc|cc|cc}
\hline
\hline
Deformation&\multicolumn{4}{c}{Lower resonance}&\multicolumn{4}{|c}{Upper resonance}&\multicolumn{2}{|c}{Total}&\multicolumn{2}{|c}{Sum rule}\\
\hline
$\delta$&$\omega_{\rm SR,1}$&$\sigma_{\rm SR,1}$&$\Gamma_{\rm SR,1}$& $B_{\rm SR,1}$ &$\omega_{\rm SR,2}$&$\sigma_{\rm SR,2}$&$\Gamma_{\rm SR,2}$& $B_{\rm SR,2}$ &$\omega_{\rm SR}$& $B_{\rm SR}$ &      $\omega_{\rm SR}$& $B_{\rm SR}$ \\
    &    (MeV) &(mb)    & (MeV)  &($\mu_N^2$)  &      (MeV)&(mb)     & (MeV)   &($\mu_N^2$)  &      (MeV)   &($\mu_N^2$)  &   (MeV)& ($\mu_N^2$) \\
\hline
0.25&  1.95(4) &~0.41(4)& 0.61(5)&   4.5(6)    &  2.48(6)  &0.49(6)  &0.90(10) & 6.3(10)     &  2.26(5)     & 10.8(12)    &   2.2  &9.9      \\
\hline
\hline
\end{tabular}
\\
\label{tab:strengths}
\end{table*}

During the last decades several SR models have been launched to explain the results
of the ($\gamma$, ${\gamma} ^{\prime}$) and ($e, e^{\prime}$) reactions~\cite{heyde2011}.
Very recent theoretical work on the scissors mode by Balbutsev, Molodtsova, and Schuck~\cite{balbutsev2013} 
postulates a
new additional mode, the isovector spin scissors mode, that may explain the appearent splitting
of the scissors structure. However, the results of these calculations
are rather qualitative at the present stage as pairing correlations are not taken into account. 
Furthermore, an important challenge is to
explain why the splitting appears in the actinides and not in the rare-earth region.

In this work we have chosen the sum-rule approach~\cite{lipparini1989}, which 
is a rather fundamental way to predict both $\omega_{\rm SR}$ and $B_{\rm SR}$ consistently.
We follow the description of Enders {\em et al.}~\cite{enders2005}
with the exception that the ground-state moment of inertia
will be replaced by the rigid-body moment of inertia. The outline for the 
quasi-continuum was recently presented~\cite{gsf2014}, and we only give
a summary of the formulas here.

The inversely and linearly energy-weighted sum rules are given by~\cite{gsf2014}
\begin{eqnarray}
S_{+1}&=&\frac{3}{2\pi}\Theta_{\rm rigid}\delta^2\omega_D^2\left(\frac{Z}{A}\right)^2\xi ~\left[ \mu^2_N {\rm MeV} \right], \\
S_{-1}&=&\frac{3}{16\pi}\Theta_{\rm rigid}\left(\frac{2Z}{A}\right)^2 ~\left[ \mu^2_N {\rm MeV}^{-1} \right].
\end{eqnarray}
The two sum rules can now be utilized to extract the SR centroid and strength:
\begin{eqnarray}
\omega_{\rm SR}&=& \sqrt{S_{+1}/S_{-1}} \nonumber\\
       &=&|\delta| \omega_D\sqrt{2\xi},  \label{eq:omega} \\
B_{\rm SR}&=& \sqrt{S_{+1}S_{-1}} \nonumber \\
          &=&\frac{3}{4\pi}\left(\frac{Z}{A}\right)^2 \Theta_{\rm rigid}|\delta| \omega_D\sqrt{2\xi} \nonumber \\
          &=&\frac{3}{4\pi}\left(\frac{Z}{A}\right)^2 \Theta_{\rm rigid}\omega_{\rm SR}.
          \label{eq:b}
\end{eqnarray}
The rigid-body moment of inertia is taken as
\begin{equation}
    \Theta_{\rm rigid} =\frac{2}{5}m_N r_0^2 A^{5/3}(1+0.31\delta),
    \label{eq:theta}
\end{equation}
    with $r_0=1.15$ fm and $\delta$ is the nuclear quadrupole 
    deformation\footnote{The quadrupole deformation  parameter $\delta$
    relates to lowest order to $\epsilon_2$ and $\beta_2$ as
    $\delta \approx \epsilon_2 \approx \beta_2\sqrt{45/16\pi}$.} taken from~\cite{goriely2009}.
The reduction factor
\begin{equation}
\xi=\frac{\omega_Q^2}{\omega_Q^2 + 2\omega_D^2}
\end{equation}
depends on the IVGDR and ISGQR frequencies of
\begin{eqnarray}
\omega_D &\approx& (31.2A^{-1/3} + 20.6A^{-1/6})(1-0.61\delta) {\rm MeV},
\label{eq:wd} \\
\omega_Q &\approx& 64.7A^{-1/3}(1-0.3\delta) {\rm MeV}.
\end{eqnarray}
The location of the IVGDR from systematics [Eq.~(\ref{eq:wd})] gives $\omega_D= 11.3$~MeV. However,
the GEDR structures of Fig.~\ref{fig:GDR} have clearly a higher average centroid. From the GEDR resonance parameters
of Table~\ref{tab:GDR_param} we find $\omega_D= 13.4$~MeV, which we adopt for the sum-rule estimates.

The two last columns of Table~\ref{tab:strengths} show the predicted $\omega_{\rm SR}$ and $B_{\rm SR}$ 
from the sum-rule estimates. Both values are in excellent agreement with our measurements.

 \begin{figure}[h]
 \begin{center}
 \includegraphics[clip,width=\columnwidth]{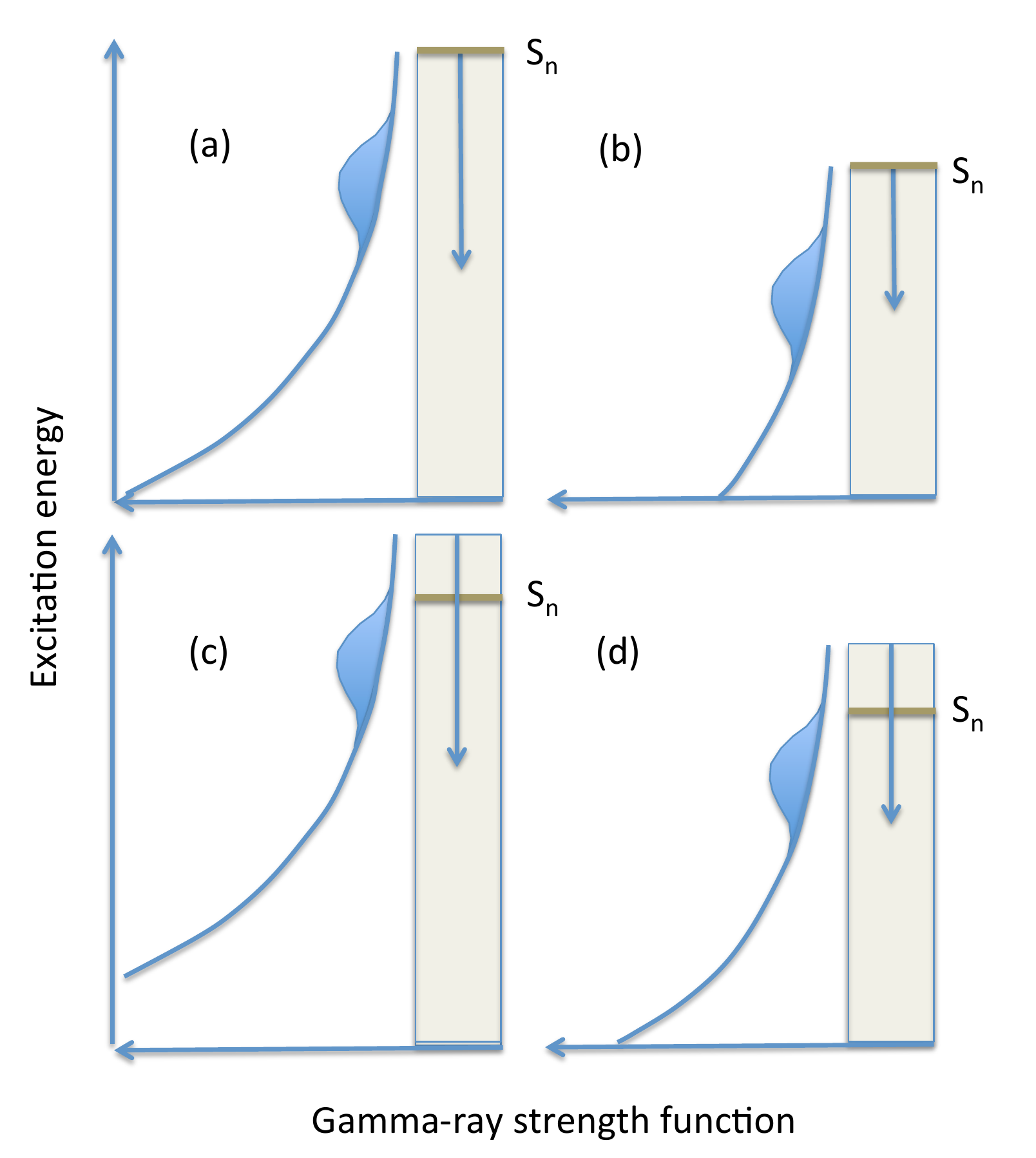}
 \caption{(Color online) Schematic view of how the SR (blue bump) influences the $\gamma$-decay rates. 
 The scenarios are: (a) $E\approx S_n$ and high $S_n$,
 (b) $E\approx S_n$ and low $S_n$, (c) $E > S_n$ and high $S_n$, and (d) $E > S_n$ and low $S_n$.
 If the centroid of the $\gamma$ energies (arrow) overlaps with the centroid of the SR, the
 $\gamma$ branch may increase significantly (up to a factor of two). The
 $^{238}$Np nucleus with a relative high separation energy of $S_n=5.488$~MeV, corresponds to case (a).
 With neutron energies of several MeV, the influence of the SR will diminish.
 }
 \label{fig:cases}
 \end{center}
 \end{figure}

\section{Calculations of the ($n,\gamma$) cross section}

The $\gamma$SF in the quasi-continuum is the
quantity that directly relates to the reaction rates in e.g.~astrophysical environments. 
For example for the r-process,
which involves nuclei with extreme $N/Z$ ratios, 
the decrease in neutron-separation energy with neutron number is expected to give an
increasing impact from the SR on the reaction rates.
The SR represents also an important ingredient for the simulations of fuel cycles for
fast nuclear reactors. 

In Fig.~\ref{fig:cases} the influence of the SR is schematically shown for four cases.
It is obvious that if the initial state "see" much of the high-energy tail of the
$\gamma$SF, the low-lying SR strength will have less importance. This happens 
in panels (a) and (c). The higher overlap of the SR with the first-generations $\gamma$s appears
in cases (b) and (d).
In $^{238}$Np the binding energy is relatively high with $S_n=5.488$~MeV [case (a)],
which means that only the high-energy part of the SR strength distribution comes into play.

In order to study the impact of the SR for $^{238}$Np, we have performed calculations of 
the $(n,\gamma)$ cross section with the TALYS code~\cite{koning2008}. 
Experimental $(n,\gamma)$ cross sections are rather well known for $^{238}$Np, making this a good test ground for such calculations.
In particular, a recent experiment at the DANCE facility~\cite{esch2008} has provided data with small statistical errors 
for incoming neutron energies up to $\approx 300$ keV.
    
For the TALYS input we have used functions that decribe the observed level density 
and $\gamma$SF (data from Figs.~\ref{fig:rhotot} and \ref{fig:GDR}, respectively). 
For the neutron optical-model potential, we have used the 
global parameterization of Koning and Delaroche~\cite{koning2003}, but with adjusted values for the parameter $a_V$  
using a scaling factor of $0.65$ to obtain agreement with the 
evaluated $s-$wave neutron strength function of $S_0 = 1.02(6)\times 10^{-4}$~\cite{mughabghab2006}.
 \begin{figure}[t]
 \begin{center}
 \includegraphics[clip,width=\columnwidth]{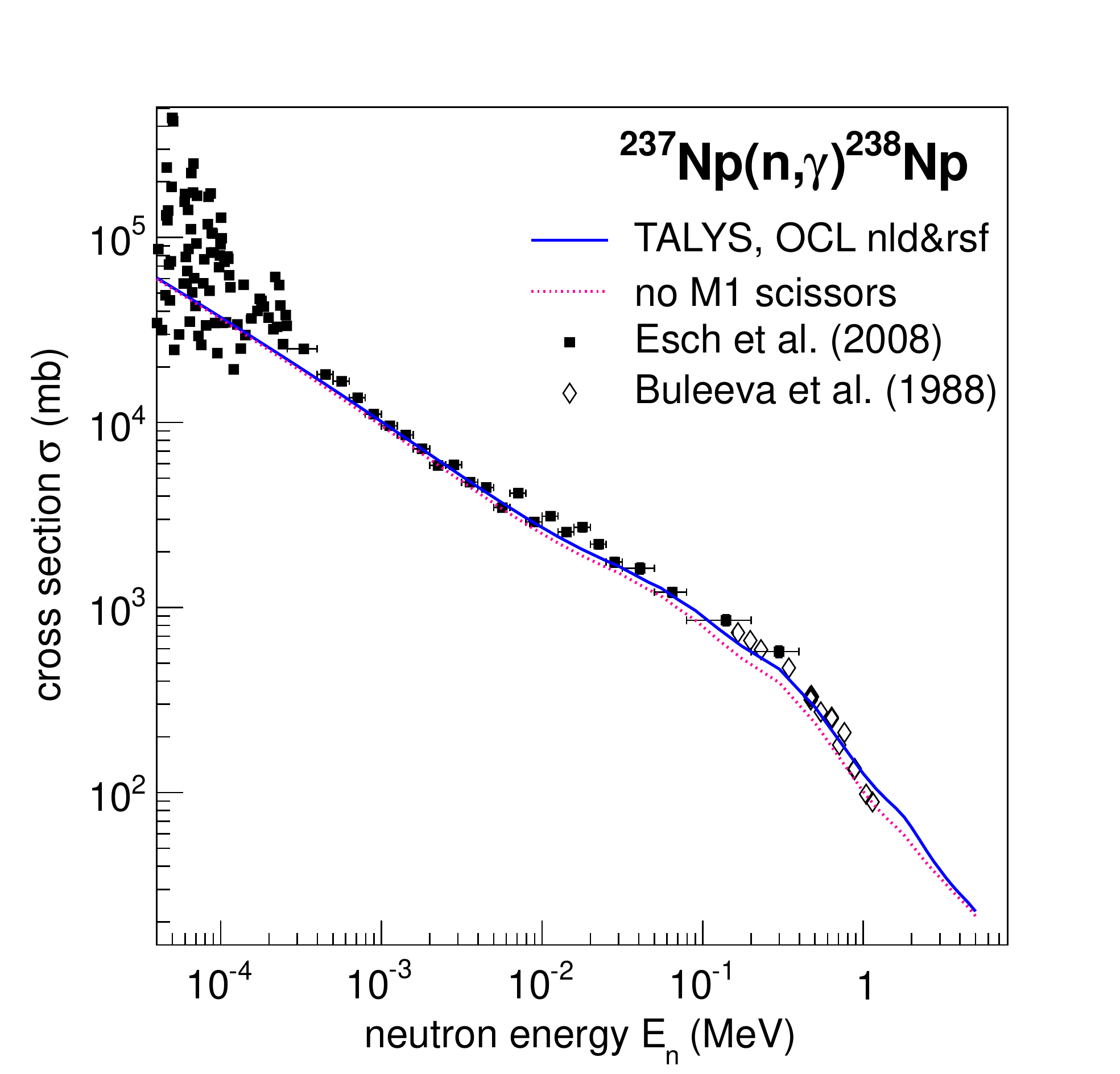}
 \caption{(Color online) Calculated $^{237}$Np$(n, \gamma)^{238}$Np cross section using
 level density and $\gamma$SF models in accordance with the data of this work. The predictions including the $M1$ scissors mode 
 (blue curve) and without (dashed line) are compared with measured data
from Esch \textit{et al.}~\cite{esch2008} (black squares) and Buleeva \textit{et al.}~\cite{buleeva1988} (open diamonds).}
 \label{fig:cross}
 \end{center}
 \end{figure}

Figure~\ref{fig:cross} shows the results of the cross-section calculations. 
The TALYS output (blue curve) is in excellent agreement with the experimentally
measured $(n, \gamma)$ cross sections from Refs.~\cite{esch2008,buleeva1988}. 
The agreements for all neutron energies above the resonance region of $E_n \geq 300$ eV give confidence to
the observed $\gamma$SF as well as the level density.
The increase in cross section due to the SR reaches a maximum of  $\approx 25$ \% for 1-MeV incoming neutrons. 
The reason for the rather small influence of the $M1$ scissors resonance on the 
$(n,\gamma)$ cross section for this case is discussed in connection with Fig.~\ref{fig:cases}; the inclusion of the SR has less impact 
because the high-energy part of the $\gamma$SF dominates the $\gamma$-decay probability. For the highest neutron energies
in Fig.~\ref{fig:cross} ($E_n \ge 5$ MeV), the SR has no practical impact on the cross-section.

\section{Conclusions}
\label{sec:con}

The level density and $\gamma$SF of $^{238}$Np have been determined using the Oslo method. The level density shows
a constant-temperature behavior similar to other actinides as recently reported for $^{231-233}$Th, $^{232,233}$Pa and $^{237-239}$U~\cite{nld2013,gsf2014}.

We observe an excess in the $\gamma$SFs in the $E_{\gamma} = 1 - 4$~MeV region,
which is interpreted as the SR in the quasi-continuum. These findings are
in contradiction with the n\_TOF results from the $(n, \gamma)^{238}$Np reaction, but in agreement with expectations for the actinide region. The underlying strength of the SR has been subtracted by extrapolating the
assumed strength from the tails of other resonances; the double humped GEDR and the two pygmy resonances. 
The SR shows a splitting into two components, however the two
components are closer in energy than observed for the other actinides.
The sum-rule applied to the quasi-continuum assuming a rigid-body moment of inertia, describes very well the
centroid and strength of the SR.

The observed level density and $\gamma$SF have been used as inputs in Hauser-Feshbach calculations with the TALYS code. 
The agreement with previously measured $(n, \gamma)$ cross sections is very gratifying. The SR strength
gives a maximum increase of 25~\% on the calculated cross section for 1-MeV neutrons.

\acknowledgements
We would like to thank J.C.~M{\"{u}}ller, E.A.~Olsen, A.~Semchenkov and J.~Wikne at the 
Oslo Cyclotron Laboratory for providing the stable and high-quality deuterium 
beam during the experiment. This work was supported by the Research Council of Norway (NFR).

\vfill

\begin{thebibliography}{99}
\bibitem{ar07} M.~Arnould, S.~Goriely, and K.~Takahashi,  Phys.~Rep.~{\bf 450}, 97 (2007).
\bibitem{kaeppler2011}F.~K\"{a}ppeler {\em et al.}, Rev. of Mod. Phys. {\bf 83}, 157 (2011).
\bibitem{chadwick2011}M.B.~Chadwick {\em et al.}, Nucl. Data Sheets {\bf 112}, 2887 (2011).
\bibitem{aliberti2006}G. Aliberti, G. Palmiotti, M. Salvatores, T.K. Kim, T.A. Taiwo, M. Anitescu, 
I. Kodeli, E. Sartori, J.C. Bosq, and J. Tommasi, Annals of Nuclear Energy {\bf 33}, 700 (2006).
\bibitem{esch2008} E-.I.~Esch, R.~Reifarth, E.M.~Bond, T.A.~Bredeweg, A.~Couture, S.E.~Glover, U.~Greife, R.C.~Haight,
A.M.~Hatarik, R.~Hatarik, M.~Jandel, T.~Kawano, A.~Mertz, J.M.~O`Donnell, R.S.~Rundberg, J.M.~Schwantes,
J.L.~Ullmann, D.J.~Vieira, J.B.~Wilhelmy, J.M.~Wouters, and A.Alpizar-Vicente, Phys.\ Rev.\ C \bf 77\rm, 034309 (2008).
\bibitem{aliberti2004}G. Aliberti, G. Palmiotti, M. Salvatores, and C.G. Stenberg, Nuclear Science and Engineering, {\bf 146}, 13 (2004).
\bibitem{Schiller00}A.~Schiller {\em et al.}, Instrum.~Methods Phys.~Res.~A {\bf 447}, 498 (2000).
\bibitem{Lars11}A.C.~Larsen {\em et al.}, Phys.\ Rev.\ C \bf 83\rm, 034315 (2011).
\bibitem{guttormsen2012}M.~Guttormsen {\em et al.}, Phys.\ Rev.\ Lett. \bf 109\rm, 162503 (2012).
\bibitem{nld2013}M.~Guttormsen {\em et al.}, Phys.\ Rev.\ C \bf 88\rm, 024307 (2013).
\bibitem{gsf2014}M.~Guttormsen {\em et al.}, Phys.\ Rev.\ C \bf 89\rm, 014302 (2014).
\bibitem{ntof2011}C.~Guerrero {\em et al.}, Journal of the Korean Physical Society, {\bf 59}, 1510 (2011).
\bibitem{siri}M.~Guttormsen, A.~B\"urger, T.E.~Hansen, and N.~Lietaer, Nucl.~Instrum.~Methods Phys.~Res.~A {\bf 648}, 168 (2011).
\bibitem{CACTUS}M.~Guttormsen {\em et al.}, Phys.\ Scr.~\bf T 32\rm, 54 (1990).
\bibitem{Gutt96}M.~Guttormsen, T.~S.~Tveter, L.~Bergholt, F.~Ingebretsen, and J.~Rekstad,
Nucl.\ Instrum.\ Methods Phys.\ Res.\ A \bf 374\rm, 371 (1996).
\bibitem{Gutt87}M.~Guttormsen, T.~Rams{\o}y, and J.~Rekstad, Nucl.\ Instrum.\
Methods Phys.\ Res.\ A \bf 255\rm, 518 (1987).
\bibitem{brink} D.M.~Brink, Ph.D.~thesis, Oxford University, 1955.
\bibitem{ENSDF}Data extracted using the NNDC On-Line Data Service from the ENSDF database.
\bibitem{RIPL3}R.~Capote {\em et al.}, Reference Input Library, RIPL-2 and RIPL-3, available online at {\it http://www-nds.iaea.org/RIPL-3/}
\bibitem{GC} A.~Gilbert and A.G.W.~Cameron, Can. J. Phys. {\bf 43}, 1446 (1965).
\bibitem{egidy2}T.~von Egidy and D.~Bucurescu, Phys.\ Rev.\ C \bf 72\rm, 044311 (2005); Phys.\ Rev.\ C \bf 73\rm, 049901(E) (2006).
\bibitem{ko90}J. Kopecky and M. Uhl, Phys.~Rev.~C {\bf 41} 1941 (1990).
\bibitem{voin1}A. Voinov, M. Guttormsen, E. Melby, J. Rekstad, A. Schiller, and S. Siem, Phys.\ Rev.\ C \bf 63\rm, 044313 (2001).
\bibitem{berman1986}B.L.~Berman, J.T.~Caldwell, E.J.~Dowdy, S.S.~Dietrich, P.~Meyer, R.A.~Alvarez, Phys.\ Rev.\ C \bf 34\rm, 2201 (1988); available at {\it http://cdfe.sinp.msu.ru/services/unifsys/index.html}.
\bibitem{heil1988}R.D.~Heil, H.H.~Pitz, U.E.P.~Berg, U.~Kneissl, K.D.~Hummel, G.~Kilgus, D.~Bohle, A.~Richter,
C.~Wesselborg, P.~von Brentano, Nucl.~Phys.~A {\bf 476}, 39 (1988).
\bibitem{margraf1990}J.~Margraf, A.~Degener, H.~Friedrichs, R.D.~Heil, A.~Jung, U.~Kneissl,
S.~Lindenstruth, H.H.~Pitz, H.~Schacht, U.~Seemann, R.~Stock, C.~Wesselborg, P.~von Brentano, A.~Zilges, Phys.\ Rev.\ C \bf 42\rm, 771 (1990).
\bibitem{yevetska2010}O.~Yevetska, J.~Enders, M.~Fritzsche, P.~von Neumann-Cosel, S.~Oberstedt, A.~Richter, C.~Romig,
D.~Savran, K.~Sonnabend, Phys.\ Rev.\ C \bf 81\rm, 044309 (2010).
\bibitem{adekola2011}A.S.~Adekola, C.T.~Angell, S.L.~Hammond, A.~Hill, C.R.~Howell, 
H.J.~Karwowski, J.H.~Kelley, and E.~Kwan, Phys.\ Rev.\ C \bf 83\rm, 034615 (2011).
\bibitem{hammond2012}S.L.~Hammond, A.S.~Adekola, C.T.~Angell, H.J.~Karwowski
E.~Kwan, G.~Rusev, A.P.~Tonchev, W.~Tornow, C.R.~Howell, and J.H.~Kelley, Phys.\ Rev.\ C \bf 85\rm, 044302 (2012).
\bibitem{heyde2011}K.~Heyde, P.~von Neumann-Cosel, A.~Richter, Rev.~Mod.~Phys.~82, 2365 (2010), and references therein.
\bibitem{balbutsev2013}E.B.~Balbutsev, I.V.~Molodtsova, and P. Schuck, Phys.\ Rev.\ C \bf 88\rm, 014306 (2013).
\bibitem{lipparini1989}E.~Lipparini and S.~Stringari, Phys.~Rep.~175, 103 (1989).
\bibitem{enders2005}J.~Enders, P. von Neumann-Cosel, C. Rangacharyulu, and A. Richter, Phys.\ Rev.\ C \bf 71\rm, 014306 (2005).
\bibitem{goriely2009}S.~Goriely, N. Chamel and J.M. Pearson, Phys.\ Rev.\ Lett. \bf 102\rm, 152503 (2009).
\bibitem{koning2008}A.J.~Koning, S.~Hilaire, and M.C.~Duijvestijn, TALYS-1.0, in
{\em Proceedings of the International Conference on Nuclear Data for Science and Technology, 22–27 April 2007, Nice, France}, edited by O.~Bersillon, F.~Gunsing, E.~Bauge, R.~Jacqmin, and S.~Leray (EDP Sciences, 2008), p. 211.
\bibitem{goriely2008}S.~Goriely, S.~Hilaire, and A.J.~Koning, Astron.~Astrophys. {\bf 487}, 767 (2008).

\bibitem{koning2003} A.~J.~Koning and J.~P.~Delaroche,  Nucl. Phys. \bf A713\rm, 231 (2003).
\bibitem{mughabghab2006} S. F. Mughabghab, Atlas of Neutron Resonances, Fifth Edition, Elsevier Science (2006).
\bibitem{buleeva1988} N.N.~Buleeva, A.N.~Davletshin, O.A.~Tipunkov, S.V.~Tikhonov, and V.A.~Tolstikov, 
    Atomnaya Energiya \textbf{65}, 348 (1988).

\end{thebibliography}
\end{document}